\documentclass[twocolumn]{aastex701}

\begin{document}


\title{A Kiloparsec-Scale Stellar Cavity in the Center of Abell402-BCG May be Caused by Dynamic Interactions with an Ultramassive Black Hole}

\author[orcid=0000-0001-5226-8349]{Michael McDonald}
\affiliation{Department of Physics, Massachusetts Institute of Technology, Cambridge, MA 02139, USA}
\affiliation{MIT Kavli Institute for Astrophysics and Space Research, Cambridge, MA 02139, USA}
\email[show]{mcdonald@space.mit.edu}


\author[]{Gourav Khullar}
\affiliation{Department of Astronomy, University of Washington, Physics-Astronomy Building, Box 351580, Seattle, WA 98195-1700, USA}
\affiliation{eScience Institute, University of Washington, Physics-Astronomy Building, Box 351580, Seattle, WA 98195-1700, USA}
\affiliation{University of Pittsburgh Physics and Astronomy Dept. 3941 O'Hara St Pittsburgh, PA 15213}
\email{}

\author[]{David Lagattuta}
\affiliation{Centre for Extragalactic Astronomy, Department of Physics, Durham University, South Road, Durham DH1 3LE, UK}
\affiliation{Institute for Computational Cosmology, Department of Physics, Durham University, South Road, Durham DH1 3LE, United Kingdom}
\email{}

\author[]{Guillaume Mahler}
\affiliation{STAR Institute, Quartier Agora - All\'ee du six Ao\^ut, 19c B-4000 Li\`ege, Belgium}
\email{}


\author[]{Shashank Dattathri}
\affiliation{Department of Astronomy, Yale University, PO. Box 208101, New Haven, CT 06520-8101}
\email{}

\author[]{Jose M. Diego}
\affiliation{Instituto de F\'isica de Cantabria (CSIC-UC). Avda. Los Castros s/n. 39005 Santander, Spain}
\email{}

\author[]{Alastair C.\ Edge}
\affiliation{Centre for Extragalactic Astronomy, Department of Physics, Durham University, South Road, Durham DH1 3LE, UK}
\email{}

\author[]{Benjamin Floyd}
\affiliation{Institute of Cosmology and Gravitation, University of Portsmouth, Dennis Sciama Building, Portsmouth PO1 3FX, UK}
\email{}

\author[]{Michael D.\ Gladders}
\affiliation{Kavli Institute for Cosmological Physics, University of Chicago, 5640 South Ellis Avenue, Chicago, IL 60637, USA}
\affiliation{Department of Astronomy and Astrophysics, University of Chicago, 5640 South Ellis Avenue, Chicago, IL 60637, USA}
\email{}

\author[]{Scott A.\ Hughes}
\affiliation{Department of Physics, Massachusetts Institute of Technology, Cambridge, MA 02139, USA}
\affiliation{MIT Kavli Institute for Astrophysics and Space Research, Cambridge, MA 02139, USA}
\email{}

\author[]{Mathilde Jauzac}
\affiliation{Centre for Extragalactic Astronomy, Department of Physics, Durham University, South Road, Durham DH1 3LE, UK}
\affiliation{Institute for Computational Cosmology, Department of Physics, Durham University, South Road, Durham DH1 3LE, United Kingdom}
\affiliation{Astrophysics Research Centre, University of KwaZulu-Natal, Westville Campus, Durban 4041, South Africa}
\affiliation{School of Mathematics, Statistics \& Computer Science, University of KwaZulu-Natal, Westville Campus, Durban 4041, South Africa}
\email{}

\author[]{Nader Khonji}
\affiliation{School of Mathematics and Physics, University of Surrey, Guildford GU2 7XH, UK}
\email{}

\author[]{Gavin Leroy}
\affiliation{Institute for Computational Cosmology, Department of Physics, Durham University, South Road, Durham DH1 3LE, United Kingdom}
\email{}

\author[]{Richard Massey}
\affiliation{Institute for Computational Cosmology, Department of Physics, Durham University, South Road, Durham DH1 3LE, United Kingdom}
\email{}

\author[]{Mireia Montes}
\affiliation{Institute of Space Sciences (ICE, CSIC), Campus UAB, Carrer de Can Magrans, s/n, 08193 Barcelona, Spain}
\email{}

\author[]{Priyamvada Natarajan}
\affiliation{Department of Astronomy, Yale University, 217 Prospect Street, New Haven, CT 06511, USA}
\affiliation{Department of Physics,  Yale University, 217 Prospect Street, New Haven, CT 06511, USA}
\affiliation{Black Hole Initiative, Harvard University, 20 Garden Street, Cambridge, MA 02138, USA}
\email{}

\author[]{Michael Reefe}
\affiliation{Department of Physics, Massachusetts Institute of Technology, Cambridge, MA 02139, USA}
\email{}

\author[]{Keren Sharon}
\affiliation{Department of Astronomy, University of Michigan, 1085 S. University Ave, Ann Arbor, MI 48109, USA}
\email{}

\author[]{Frank van den Bosch}
\affiliation{Department of Astronomy, Yale University, PO. Box 208101, New Haven, CT 06520-8101}
\email{}

\author[]{Stepane Werner}
\affiliation{Centre for Extragalactic Astronomy, Department of Physics, Durham University, South Road, Durham DH1 3LE, UK}
\affiliation{Institute for Computational Cosmology, Department of Physics, Durham University, South Road, Durham DH1 3LE, United Kingdom}
\email{}

\author[]{Adi Zitrin}
\affiliation{Department of Physics, Ben-Gurion University of the Negev, P.O. Box 653, Be'er-Sheva 84105, Israel}
\email{}

\begin{abstract}


We present new observations from \emph{JWST} that reveal a striking kpc-wide cavity in the stellar distribution of the central galaxy in the cluster Abell\,402. Supporting data from \emph{HST} allow us to rule out extinction due to dust as an explanation and, instead, suggest that this is a localized depression in the stellar density field corresponding to $\sim$$2\times10^9$ M$_{\odot}$ in missing stars within a volume of 0.5\,kpc$^3$. On larger scales, both the \emph{JWST} and \emph{HST} data show evidence for a 2.2\,kpc flattened core in the stellar distribution (on which the smaller-scale cavity is superimposed), which implies the presence of a central ultra-massive black hole with M$_{BH} = 6 \pm 4 \times 10^{10}$\,M$_{\odot}$. We report evidence for a mid-IR-bright point source at one edge of the cavity, suggesting that this black hole is actively accreting. MUSE spectroscopy reveal that this source is a LINER AGN and that there is a second candidate AGN on the opposite side of the cavity with a  relative velocity of 370\,km/s -- if real, this implies the presence of a kpc-separation dual AGN with a total binary mass of $6 \pm 2\times10^{10}$\,M$_{\odot}$, which would make this the most massive binary black hole system discovered to date. We propose that this unique stellar cavity is the result of a short-lived dynamical interaction between at least one supermassive black hole and the background stellar density field, caused either by three-body scattering during binary hardening or the induction of a dipole instability in the stellar density field.
\\
\end{abstract}

\keywords{}


\section{Introduction}

In the centers of the most massive galaxies in the Universe, so-called ``giant elliptical galaxies'', light profiles are seen to flatten out to a constant surface brightness ``core'' region \citep{lauer95,lauer05}.
The existence of a core is not naturally predicted by current galaxy formation models for massive galaxies in a $\Lambda$CDM universe \citep{nfw}, therefore this suggests that a secondary mechanism is responsible for removing stars from the high density innermost regions. The most likely culprit, given the mass of the evacuated stars, has been proposed to be the on-going merger of a pair of SMBHs \citep{milo02,volonteri03,boylan04,graham04,merritt06,rantala18,khonji24}.
According to the standard paradigm for structure formation, massive galaxies form primarily via mergers with smaller galaxies. Given that most, if not all, galaxies appear to harbor a central SMBH, the merger of black holes is hence anticipated to occur frequently as well. In particular, at late times, massive galaxies at the centers of clusters are expected to be the site of active merging events \citep{dubinski98,delucia07}.


For two SMBHs to merge, they first need to shed sufficient orbital angular momentum to shrink their orbit to sufficiently small separation such that gravitational wave emission can carry away the remaining orbital energy in less than the age of the Universe. 
While such gravitational waves from a merging supermassive black hole binary are yet to be detected, recently NANOGraV and other pulsar timing array experiments have reported evidence for the existence of a stochastic gravitational wave background from
the collective merging of binary black holes in galactic nuclei \citep{nanograv23}.
It is believed that the initial stage of this process, before gravitational waves become important, involves three-body scattering with stars, which would lead to the ``scouring'' of the stars and dark matter in the inner $\sim$kpc of the most massive galaxies \citep{frigo21,nasim21,khonji24}. This deficit of stars can be further enhanced if the merger induces a recoil on the remnant, which would remove the SMBH from the center of the gravitational potential and lead to a rapid expansion of stellar orbits behind it \citep{boylan04,favata04,nasim21,khonji24}.

Such large, diffuse cores in the stellar distribution have been observed in several massive galaxies \citep{postman12,bonfini16,thomas16,rantala18,mehrgan19}, and it has been argued that the physical size of the observed core can be directly related to the mass of the black hole that likely created it. Indeed, the masses of some of the most massive SMBHs have been deduced from the sizes of the core regions carved out by them in their host galaxy light profiles \citep{postman12,dullo19}. This relation hinges on the idea that the core was formed via dynamical interactions with a binary SMBH, either during or after the merger.
Despite significant observational effort expended studying cores in massive galaxies and on the theory side predicting mechanisms for their formation, there remains little direct observational support for this picture -- the only system for which a binary SMBH has been directly detected to be altering the stellar distribution is NGC\,5419 \citep{neureiter23}. In this system, two closely-separated point sources appear to be driving kinematic disturbances in the stellar distribution, which will eventually lead to the formation of a flattened core. In addition, these large, flat stellar cores may be dynamically unstable, as shown by \cite{dattathri25a}, leading to the development of a long-lasting dipole in the stellar distribution. This dipole arises from the sharp transition between the flat inner core and steep outer profile, and could lead to off-center central supermassive black holes due to the interaction of the black hole and the stellar dipole. Such off-center black holes have been observed in a variety of lower-mass systems \citep[e.g.,][]{binggeli00,reines20,mezcua24}. Given that stellar dipoles have not been observed in any of the known flat-core systems, it may that they are too weak to detect or that there is a mechanism to suppress them.

Here, we present observations of the central galaxy in Abell\,402, which is a massive galaxy cluster at $z=0.322$.  This work makes use of new imaging data from the \emph{JWST} Near Infrared Camera (NIRCam), which reveals a kpc-scale cavity embedded in a large diffuse core at the center of the galaxy. This cavity was previously reported by \cite{repp18}, based on \emph{HST} observations, and was speculated to be a patch of dust near the galaxy center. With these new data, we present compelling evidence that this is, in fact, an absence of stars, and argue that it is most likely due to dynamical interactions with a merging pair of SMBHs. The paper is structured as follows: in \S2, we present the data that is used in this work; in \S3 we present a photometric analysis of the stellar cavity; in \S4 we present a photometric analysis of the diffuse core and a central supermassive black hole estimate, in \S5 we present a spectroscopic analysis of two candidate AGN on either side of the cavity; and in \S6 we present three possible scenarios for the formation of the stellar cavity that are consistent with the observations.

Throughout this work, we assume a $\Lambda$CDM cosmology, with H$_0$ = 70 km s$^{-1}$ Mpc$^{-1}$, $\Omega_M=0.3$, and $\Omega_{\Lambda}=0.7$. All error bars are 1$\sigma$ confidence intervals, unless explicitly stated otherwise.

\begin{figure*}[htb]
\includegraphics[width=0.99\linewidth]{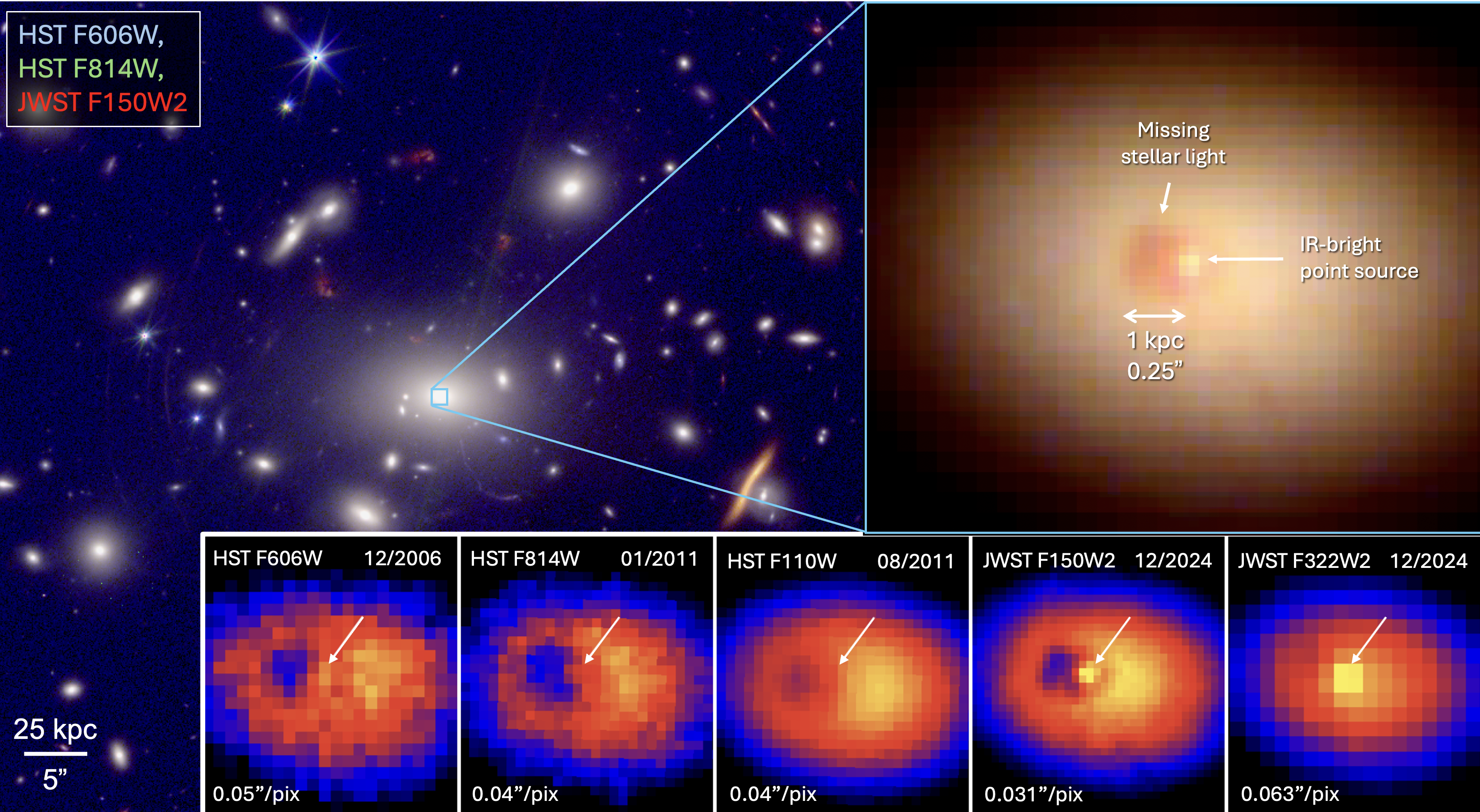}
\caption{{\bf Multi-wavelength observations of the central galaxy in Abell 402.} \emph{Left:} Three-color image of the cluster core based on data from \emph{HST} and \emph{JWST} (north is up, east is left). \emph{Upper right:} Zoom-in on the inner $2^{\prime\prime}\times 2^{\prime\prime}$ of the central galaxy in Abell\,402. This zoomed image reveals a 0.25$^{\prime\prime}$-wide cavity in surface brightness, as well as a bright point source on the western edge of the depression. \emph{Lower right:} Images of the inner region of the central galaxy in five different filters at four different epochs. 
The point source (depicted with a white pointer in all panels) is brightest in the redder JWST band, where the combination of the bright point source and degraded PSF obscures the presence of any cavity.}
\label{fig:fig1}
\end{figure*}

\section{Data}
This paper makes use of new and archival data from a variety of telescopes, which we describe below. We limit this section to the description of the data acquisition and basic reductions, and will discuss the details of our analyses in subsequent sections where appropriate. We also defer discussion of supporting multiwavelength data in the radio and X-ray to the discussion at the end of this paper, once their motivation is clear.

\subsection{Optical/IR Imaging: JWST and HST}

This work makes use of new imaging data from the \emph{JWST} Near Infrared Camera (NIRCam) at short (F150W2) and long (F322W2) wavelengths from the JWST SLICE (Strong Lensing and Cluster Evolution) program (\#5594; PI G.\ Mahler). These data were reduced using a custom pipeline developed for these shallow survey data, which is described in full detail in other papers \citep{rigby25,cerny25}. We briefly describe the JWST data reduction process here. We utilize Level 1b products from MAST using an STScI-adapted custom Python script. For data reduction, we employ scripts modified from the default JWST pipeline (v1.15.1), and the CRDS calibration reference data system pipeline mapping (CRDS, pmap) 1303. We process Level 2A data products to apply a custom de-striping algorithm to correct for 1/f noise and jumps between amplifiers. We then run the de-striped images through the same modified pipeline, to generate Level 3 F150W2 and F322W2 dithered science-ready mosaics. The filters are WCS-matched to \emph{Gaia} as per the specifications of the default pipeline. For more details on the various steps and systematics involved in this process, please refer to the complete description of the custom pipeline \citep{rigby25}.


We supplement these near-infrared imaging observations with archival imaging data from the \emph{HST} Advanced Camera for Surveys (ACS) and Wide Field Camera 3 (WFC3) in the F606W, F814W, and F110W filters, yielding high angular resolution 5-band imaging spanning 0.5--4$\mu$m over the cluster core.  These data were obtained with the WFC-ACS camera (F606W, F814W) and with the WFC3-IR camera (F110W), and were previously published by \cite{repp18}. We reduce these data using the standard, automated MAST pipeline, which is sufficient for the purpose of this study (aperture photometry and morphological comparison between bands).

\subsection{Optical Spectroscopy: MUSE}
MUSE observations of Abell\,402 were obtained between September 2017 and September 2019, as part of the MUSE Guaranteed Time Observations (GTO) Lensing Clusters program (PI J.\ Richard). The full data set consists of 30 1000-second exposures, all centered around the cluster BCG. All data were taken in Wide-Field mode using the nominal (N) wavelength range. However, the first six exposures were acquired without ground layer adaptive optics (WFM-NOAO-N mode), while all subsequent exposures used the AO system (WFM-AO-N mode). However, due to the excellent conditions on these nights, the ground layer correction was minimal and there was virtually no difference in the delivered image quality between these observations. As such, we combine all observations in our analysis. These data were reduced following standard procedures as described in previous works \citep{richard2021,lagattuta2022}.

Spectral modeling was performed in narrow wavelength intervals around emission lines. We choose a relatively featureless and emission line-free region of the central galaxy to extract a continuum spectrum, and then include the normalization of this continuum as a free parameter in our models. We combine this one-parameter continuum model with a series of gaussians to represent emission lines. We fit each emission line independently over a narrow ($\sim$50\AA) interval, allowing the line position, width, and flux to vary independently for each line.  In general, the observed emission lines are quite high signal-to-noise and are, thus, relatively insensitive to choices in methodology. Where errors are quoted these represent bootstrapping errors over the measured uncertainties, continuum models, and initial conditions.

\section{A Kiloparsec-Wide Cavity in the Stellar Distribution of Abell 402 BCG}

Figure \ref{fig:fig1} shows the central galaxy in Abell\,402, as observed by HST and JWST. At the very center of this galaxy are two distinct features: a bright point-like source at the galaxy center, and a kpc-wide dark region to the east of this source. The point source is brightest in the redder NIRCam band (F322W2) and has a strong wavelength dependence characteristic of an active galactic nucleus (AGN).
The dark region was first identified in observations with HST, and was speculated to be a patch of dust near the center of the galaxy \citep{repp18}. However, the fact that this feature is still present in near-infrared data from JWST -- which should be largely unaffected by dust -- presents a major challenge for this interpretation.  In the Appendix, we demonstrate that the cavity and point source are present and consistent in position and morphology at all wavelengths, and that the subtle differences between these five panels can be attributed to a strongly wavelength-dependent point source and a varying PSF across the five bands, leading to an apparent filling in of the cavity at F110W and F322W2.

\begin{figure*}[t]
\includegraphics[width=0.99\linewidth]{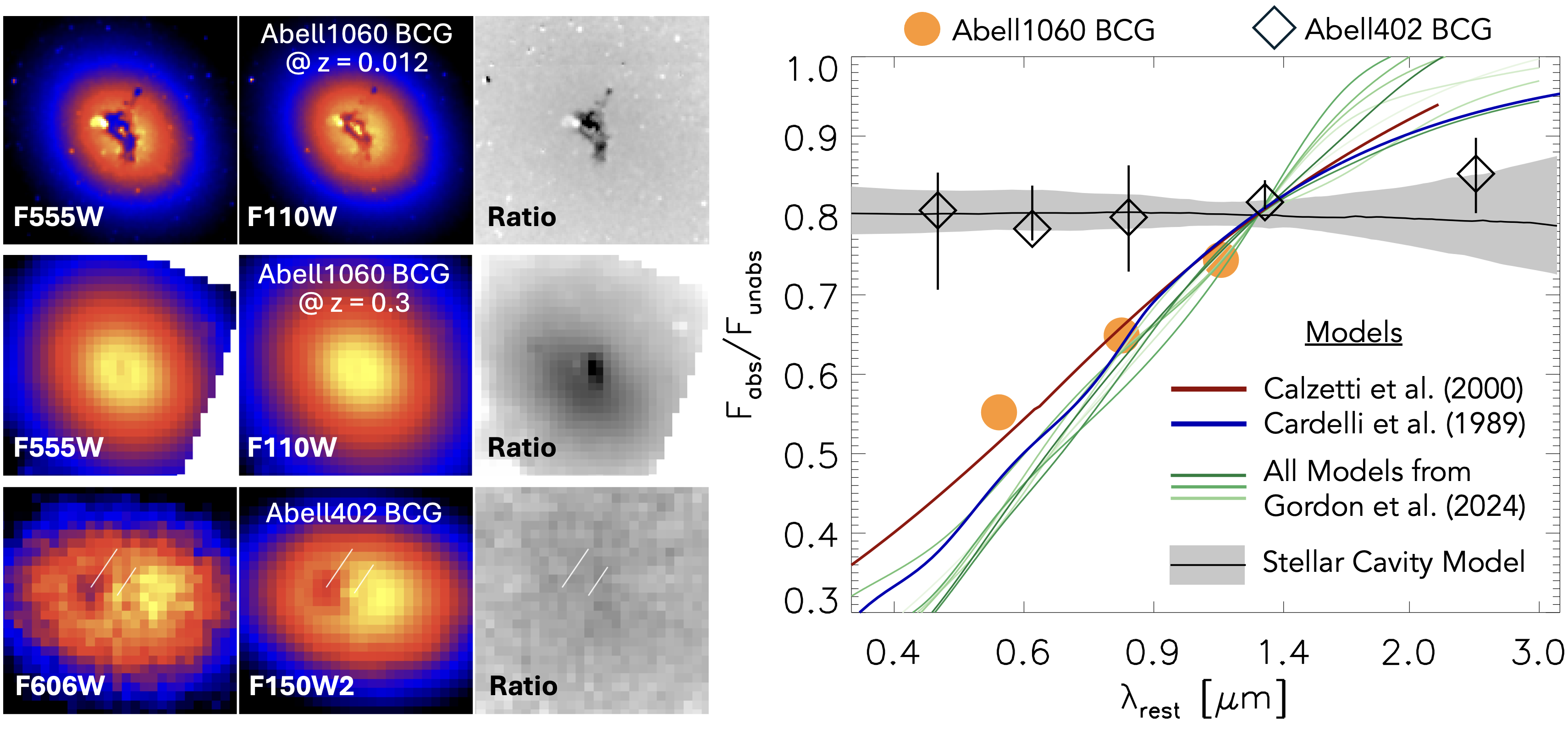}
\caption{\emph{Left:} Comparing the dark patch in Abell1060-BCG and Abell402-BCG. In the upper row, we show F555W and F110W observations of the central galaxy in Abell1060 ($z=0.012$), which has a well-known ring of dust in the inner kpc. The F555W image has been smoothed and rebinned to match the F110W image. In the right column we show the ratio of these two images, which highlights the wavelength-dependent intensity of the dust ring. In the middle row, we have simulated how Abell1060 would appear at $z=0.3$ by adjusting the binning/smoothing based on the larger angular diameter distance -- the dust ring now appears to be an unresolved cavity in the galaxy center, similar to what is observed in Abell402. The dust feature is still highly significant in the ratio image. In the bottom panel, we show Abell402, where we have smoothed and rebinned the F150W2 data to match the F606W image. In this case, the ratio image contains no structure, suggesting that the observed cavity is missing stars, rather than dust. \emph{Right:} Intensity of the dark patch compared to the expectation value from a symmetric model, as a function of wavelength for Abell402-BCG  (diamonds) and Abell1060-BCG (orange circles). This figure shows that the constant depth of the cavity in Abell402 with wavelength (dashed line) is inconsistent with predictions from extinction models. We compare to the standard models for the Milky Way \citep[blue;][]{cardelli89} and for the starburst galaxies \citep[red;][]{calzetti00}, but also compare to eight models that span different regions in the LMC, SMC, Milky Way, M31, and M33 \citep[green;][]{gordon24}. All of these models predict a factor of $\sim$4 difference in rest-frame extinction between the F150W2 and F606W bands. In contrast, the feature in Abell1060-BCG is fully consistent with dust. We also compare to a simple model in which we remove a cylindrical volume of stars from the midplane of the galaxy, with spectral shape based on the measured color gradient of the galaxy -- this model (in grey) describes the data excellently.
}
\label{fig:fig2}
\end{figure*}

In Figure \ref{fig:fig2} we contrast the central galaxy in Abell\,402 with the central galaxy in the nearby cluster Abell\,1060 ($z=0.012$), which has a well-known dust ring at its center \citep{laine03}. When we simulate the appearance of Abell\,1060 at $z=0.3$ (the redshift of Abell\,402), the dust ring appears as an unresolved ``cavity'' in the galaxy center, similar to Abell\,402. However, in Abell\,402, the integrated intensity of the dark patch is independent of wavelength.  Dust models \citep[e.g.,][]{cardelli89,fitzpatrick99,calzetti00} predict that, given the depth of the feature in the F150W2 band, the F606W band should have a factor of 4 times more absorption than is observed -- this is indeed what we observe in Abell\,1060. This is highlighted in the right panel of Figure \ref{fig:fig2}, where we estimate the amount of missing flux in the dark patches of Abell\,402 and Abell\,1060 as a function of wavelength. This figure makes it clear that the feature in Abell\,402 is inconsistent with dust models.
Instead, the data is well-fit by a model that mimics the removal of stars within 0.5\,kpc of the midplane of the galaxy by taking the ratio of the projected light along two different paths (see Appendix A.1 for more details).
We note that this feature cannot be produced by gas, which would require ionized gas densities that are unrealistically high ($n_e \sim 100$ cm$^{-3}$). Likewise, it could not be produced by nearby macroscopic absorbers (e.g., asteroids), which would move on the sky on timescales of decades (the cavity was first observed in 2006), or distant macroscopic absorbers, for which one would need to invoke unrealistic distributions of solid objects that violate the mass budget of the universe

Having ruled that absorption due to dust is inconsistent with the data, we conclude that the dark region in Figure \ref{fig:fig1} is a cavity in the stellar distribution. To estimate the missing stellar mass in the kpc-wide cavity, we utilize the image decomposition described in Appendix A.1, which yields an analytic model for the cavity as a function of wavelength. We fit the spectral energy distribution of the cavity as described in Appendix A.2, finding 
that the total stellar mass that is contained in this negative-intensity component is $M_{*,cavity} = 2.1 \pm 0.9 \times 10^9$ M$_{\odot}$, which represents $<$1\% of the total stellar mass of the galaxy. 
%
This estimate does not include the mass in dark matter that would be associated with these missing stars, which may account for an additional $\sim10^9-10^{10}$\,M$_{\odot}$, depending on the underlying shape of the dark matter profile. 

Removing such a large, concentrated mass in stars and dark matter requires dynamical interactions with a scatterer of roughly equal or greater mass. As described in \S1, dynamical interactions can produce such a signature in three distinct ways: (i) three-body interactions during the ``hardening'' of the SMBH binary would scatter nearby stars to either highly-eccentric or unbound orbits \citep{rantala18,frigo21,nasim21,khonji24}, (ii) a stalled SMBH binary in a flat stellar density field can induce a dipole instability in the distribution of stars which can appear as a cavity for certain geometries \citep{dattathri25a,dattathri25b,vandenbosh25}, or (iii) the recent merger of two SMBHs could induce a recoil in the remnant, effectively removing the SMBH from the center of the gravitational potential and leading to a rapid expansion of stellar orbits within its (now evacuated) sphere of influence \citep{boylan04,favata04,nasim21,khonji24} or, if the SMBH does not reach the escape velocity, inducing a sloshing motion in the core which can transfer energy and angular momentum to stars in the core. To determine which, if any, of these scenarios are realistic for Abell\,402, we must first establish the presence and infer the properties of any central SMBHs.

\section{A Large Core Indicating the Presence of an Ultramassive Black Hole}

The presence of a large, diffuse core in the center of a massive galaxy is often used as direct evidence for the presence of a supermassive black hole \citep[e.g.,][]{postman12,bonfini16,thomas16,rantala18,mehrgan19}. Assuming that these cores are formed via scouring by merging supermassive black holes, the extent of the core (i.e., the ``break radius'') can be used as a proxy for black hole mass \citep[e.g.,][]{lauer07,kormendy09,dullo19}. When performing the two-dimensional modeling of the central galaxy to extract the properties of the cavity and central point source (see Appendix A.1), we found that the best-fitting model was the ``Nuker'' profile \citep{lauer95}, with a major axis break radius at F150W2 of 15.3 pixels (0.48$^{\prime\prime}$ or 2.2\,kpc). This flat-core model, when combined with a point source and a cavity (see Appendix A.1), yields an excellent fit to the data with no strong residuals. This break radius is amongst the largest ever measured, with the largest core radii measured to date being 2.7\,kpc and 3.8\,kpc in Abell\,2261 BCG \citep{dullo19} and Abell\,2029 BCG \citep[IC\,1101; ][]{dullo17}, respectively. Based on a scaling relation between the break radius and the black hole mass for 11 cored ellipticals with direct M$_{BH}$ measurements \citep{dullo19}, we estimate a black hole mass of $6\pm4 \times 10^{10}$\,M$_{\odot}$, where the uncertainty comes from the uncertainty in the scaling relation..

\begin{figure}
\includegraphics[width=0.99\linewidth]{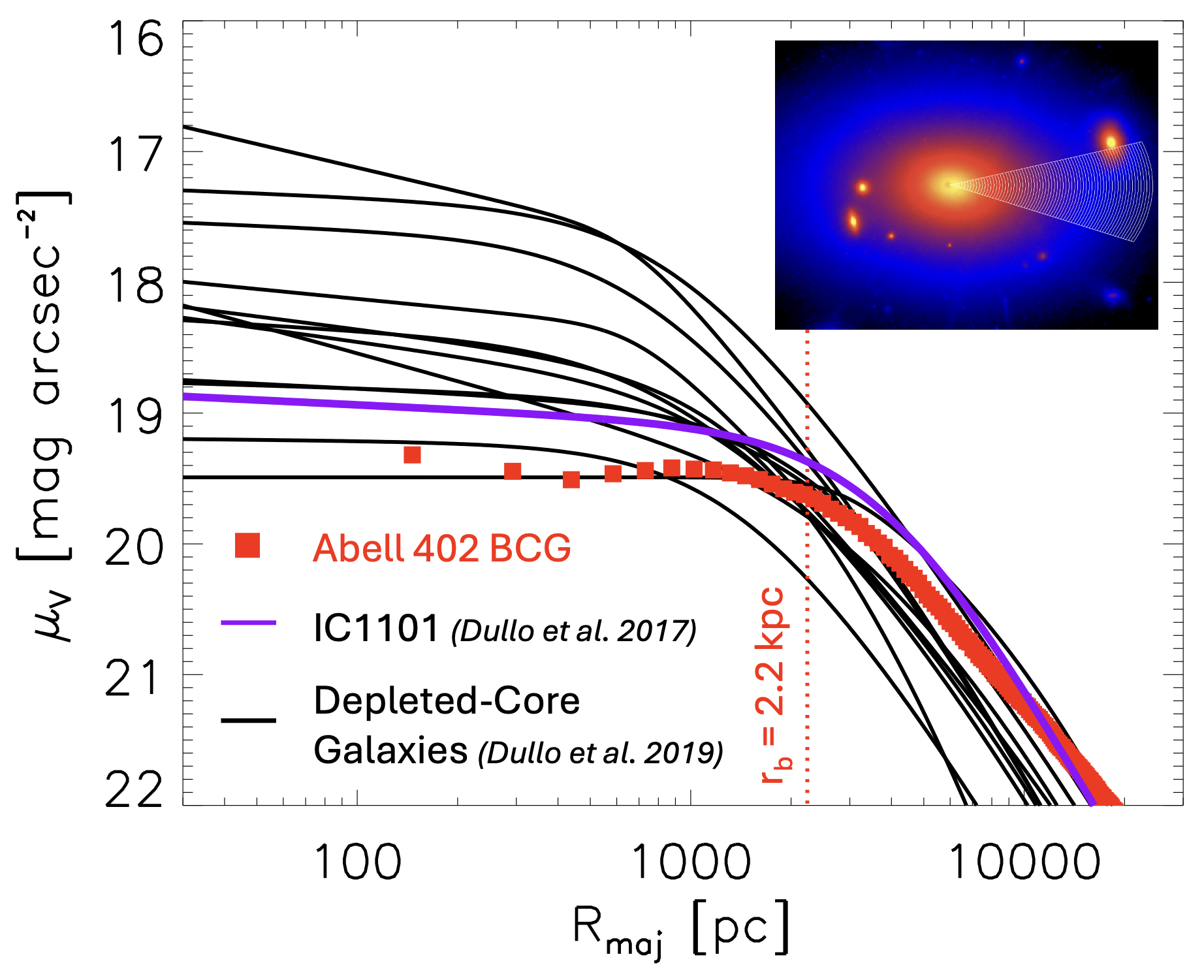}
\caption{This figure shows, in red, the F150W2 surface brightness profile extracted from the central galaxy in Abell\,402 BCG along a wedge as shown in the inset. For comparison, we show the galaxy with the largest core measured to date, IC\,1101 \citep{dullo17}, and a sample of cored galaxies from \cite{dullo19}. All profiles have been corrected for cosmological dimming and k-corrected to the rest-frame V band assuming an old stellar population. The vertical red dotted line shows the location of the break radius measured from the two-dimensional fit described in Appendix A.1. This figure highlights the similarities between IC\,1101 and Abell\,402 BCG, and suggest that Abell\,402 BCG likely harbors an ultramassive black hole.}
\label{fig:sbprofile}
\end{figure}

Rather than relying on a model fit to the imaging data, we can also consider a one-dimensional surface brightness profile to assess the degree to which the central light distribution is flattened. In Figure \ref{fig:sbprofile}, we show a one-dimensional surface brightness profile computed from an elliptical wedge centered on the central point source and with a shape matching the large-scale isophotal ellipticity (see inset in Figure \ref{fig:sbprofile}). This wedge extends to the west, avoiding the stellar cavity entirely. We correct the measured surface brightness profile in the F150W2 band for cosmological dimming ($\propto(1+z)^4$), and convert from the F150W2 band in the observed frame to the V band in the rest frame, assuming a 10\,Gyr old simple stellar population. This allows us to directly compare to profiles from \cite{dullo17} and \cite{dullo19}, which are provided in the same bandpass. These comparison profiles comprise the most extended cores known, most of which are considerably \emph{less} cored than Abell\,402 BCG. This comparison confirms the presence of a large, diffuse core in Abell\,402 BCG, which is consistent in size and surface brightness to the exceptionally-large core detected in IC\,1101, and is clearly not a result of modeling assumptions. The combination of the two-dimensional modeling in Appendix A.1 and the one-dimensional surface brightness profile in Figure \ref{fig:sbprofile}, suggest that the central galaxy in Abell\,402 hosts an ultramassive black hole with a mass $>$10$^{10}$\,M$_{\odot}$.

\section{Evidence for a Kiloparsec-Wide Binary AGN in Abell 402 BCG}
The JWST data presented in Figure \ref{fig:fig1} shows evidence for a mid-IR-bright point-source at the center of the galaxy and on the Western edge of the cavity. After spatially and spectrally decomposing the point source and the galaxy (see Appendix A.1), this point source has a spectral energy distribution fully consistent with a mid-IR bright power law (see Appendix A.2 for more details), as is commonly seen for AGN. 
The combination of an IR-bright powerlaw continuum and the fact that the source is point-like at JWST angular resolution suggests that this source is an actively accreting SMBH. Coupled with the large observed core (described in the previous section), this emission is most likely associated with a $>$10$^{10}$\,M$_{\odot}$ central black hole.


\begin{figure*}
\includegraphics[width=0.99\linewidth]{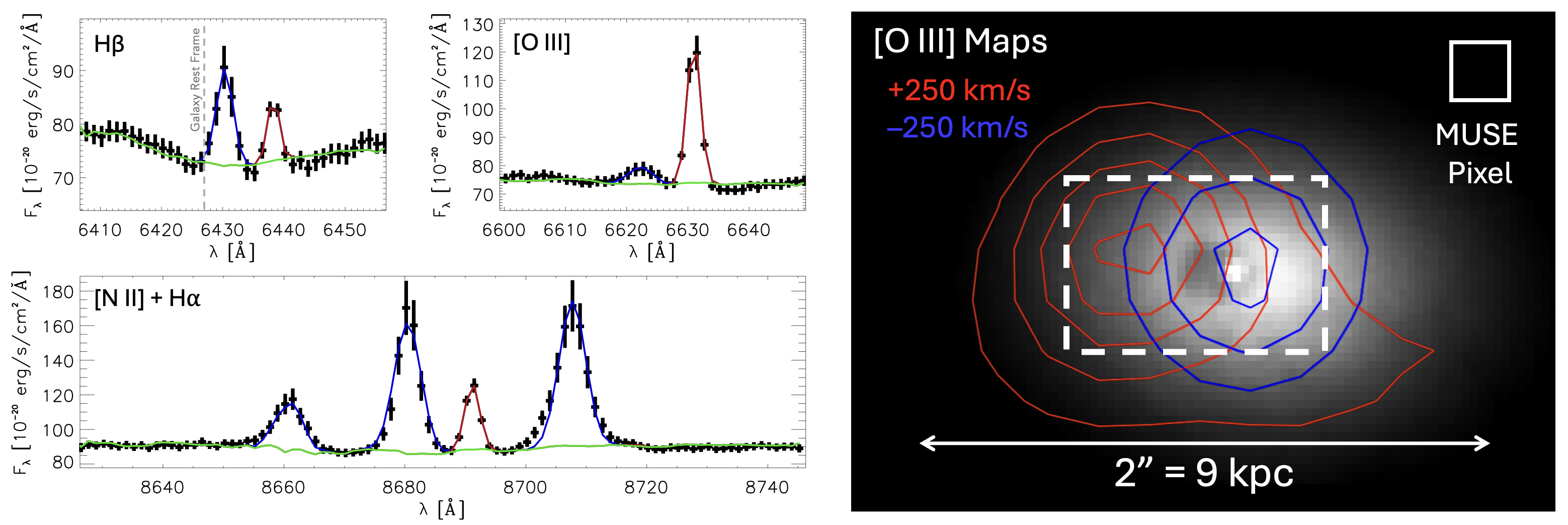}
\caption{\emph{Left:} Integrated spectra from a 1$^{\prime\prime}$-wide region centered on the stellar cavity, showing the brightness of various lines. The two velocity components are clearly offset, with the blue-shifted source having LINER-like emission line ratios (low [O\,\textsc{iii}]/H$\beta$ ratio, high [N\,\textsc{ii}]/H$\alpha$ ratio), while the red-shifted source has very bright [O\,\textsc{iii}] emission but no detectable [N\,\textsc{ii}]. In the H$\beta$ panel, the rest-frame velocity of the galaxy is shown for comparison. In all panels, the green line represents the empirical continuum model, extracted from a nearby region free of emission lines and rescaled in intensity to fit the central region. \emph{Right:} Zoomed-in image of the central galaxy in Abell\,402 in the F150W2 band. The white rectangle shows the extraction area for the MUSE spectra shown on the left. Red and blue contours show the intensity of the two kinematic components of [O\,\textsc{iii}]$\lambda$5007 shown on the left. This figure demonstrates the presence of two highly-ionized, point-like sources on either side of the stellar cavity, with high relative velocities. The eastern (red-shifted) source, while compact, appears to be embedded in a larger ($\sim$10\,kpc) blob of highly-ionized gas.}
\label{fig:fig4}
\end{figure*}

We appeal to integral field spectroscopy from MUSE to establish the presence (or absence) of a second AGN in the center of Abell\,402 BCG. In Figure \ref{fig:fig4}, we show spectra extracted from a $1^{\prime\prime} \times 0.6^{\prime\prime}$ ($5\times3$ pixels) region centered on the stellar cavity, revealing the presence of two kinematically-distinct sets of emission lines, separated along the line of sight by a velocity of 370\,km/s in the rest frame of the galaxy. Among other lines, we detect bright H$\alpha$, H$\beta$, and [O\,\textsc{iii}]$\lambda$5007 emission lines at both $z=0.3227$ and $z=0.3243$, and the [N\,\textsc{ii}]$\lambda\lambda$6548,6583 emission line doublet at $z=0.3227$ only. For reference, we measure an absorption-line redshift of the stellar component of $z=0.3225$. All of these emission lines are narrow, with no evidence for broad line AGN.

In Figure \ref{fig:fig4}, we also show the spatial distribution of the two kinematic components of the high-ionization [O\,\textsc{iii}]$\lambda$5007 line. The blue-shifted gas (which is within $\sim$100\,km/s of the stellar redshift) originates from a point-like source on the Western edge of the stellar cavity. This source is coincident with the point-like continuum source shown in Figure \ref{fig:fig1}, and shows no evidence for an extended component, further increasing the already-high likelihood that it is an AGN. The redshifted gas appears to be a point-like source on the Eastern edge of the cavity, embedded in a more extended, diffuse envelope of [O\,\textsc{iii}]-emitting gas. This extended emission may be related to a recent galaxy-galaxy merger or an ongoing dynamical interaction with the blue-shifted AGN.
The separation between the red- and blue-shifted [O\,\textsc{iii}] peaks in the plane of the sky is 2.0 $\pm$ 0.6\,kpc, which is consistent with the measured diameter of the cavity in the near-IR. Assuming that these two sources are point-like and in orbit around one another perpendicular to the plane of the sky, the implied binary mass assuming a relative velocity in the rest frame of 370\,km/s and a separation of $2.0 \pm 0.6$\,kpc is M$_{tot}$ = 6 $\pm$ 2 $\times$ 10$^{10}$\,M$_{\odot}$ (from Kepler's law), which is remarkably consistent with the estimate of the central SMBH mass based on the presence of a large, flat stellar core. The uncertainty on this estimate is dominated by the uncertainty on the separation between the two sources (Figure \ref{fig:fig4}).

\begin{figure}[t]
\centering
\includegraphics[width=0.99\linewidth]{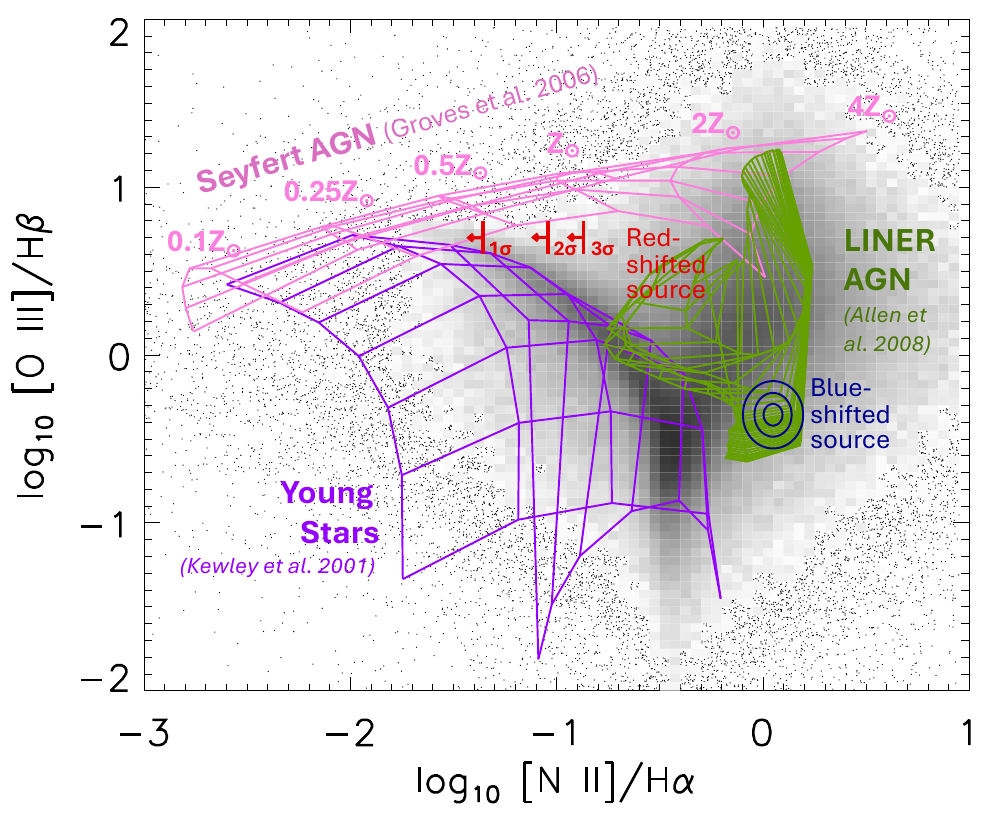}
\caption{Line ratio diagram, comparing high-ionization ([O\,\textsc{iii}]/H$\beta$) and low-ionization ([N\,\textsc{ii}]/H$\alpha$) line ratios \citep{bpt}, which separates star forming regions from active galactic nuclei and is insensitive to the effects of dust. The black/grey points in the background represent 85,224 galaxies with spectra from the Sloan Digital Sky Survey \citep{kewley06}. Where the point density becomes sufficiently high we transition to a 2D histogram. There is a dense locus of points that are consistent with predictions for H\,\textsc{ii} regions \citep{kewley01} (purple grid). The blue-shifted source on the western side of the stellar cavity has line ratios consistent with LINER AGN \citep{heckman80,allen08} (green grid) and is inconsistent with models of star formation. The red-shifted source on the eastern side of the stellar cavity has no detectable [N\,\textsc{ii}] emission, and so appears here as an upper limit in [N\,\textsc{ii}]/H$\alpha$. This source is fully consistent with models of Seyfert-like AGN with low-metallicity gas \citep{groves06} (pink grid), but is also marginally consistent with a very young and energetic starburst. We rule out the latter scenario based on the lack of blue continuum emission, which would have been clearly visible in the archival HST data at F606W.
}
\label{fig:fig7}
\end{figure}

In Figure \ref{fig:fig7}, we show the ratios of the [O\,\textsc{iii}]/H$\beta$ and [N\,\textsc{ii}]/H$\alpha$ emission lines for the red- and blue-shifted sources identified in Figure \ref{fig:fig4}. This pair of emission line ratios is commonly used to distinguish between young stellar populations, which produce low-ionization spectra, and AGN, which tend to produce both high- and low-ionization spectra \citep{bpt}. The blue-shifted source has higher [N\,\textsc{ii}]/H$\alpha$ than can be predicted by stellar photoionization, and is consistent with typical ``low-ionization nuclear emission-line regions'' (LINERs), which are thought to be arising typically from low luminosity AGN accreting at very low rates \citep{heckman80}. Combined with the presence of an IR-bright point source that exhibits a power-law SED, there is substantial evidence that the blue-shifted source is an accreting SMBH. The red-shifted source, on the other hand, is not detected at [N\,\textsc{ii}], making its spectrum challenging to interpret. The extremely-high [O\,\textsc{iii}]/H$\beta$ ratio measured for this source ($\sim$5) is at the very limit of what can be produced with the most extreme star formation, lying on the so-called ``extreme starburst line'' \citep{kewley01}. Assuming that this ionization is due to star formation yields an H$\alpha$-derived star formation rate \citep{kennicutt98} of 0.04 M$_{\odot}$ yr$^{-1}$, which would lead to a predicted surface brightness a factor of two higher than what is observed at F606W \citep{starburst99}. While ``normal'' star formation is inconsistent with the observations, we can not rule out that this eastern source is a ``green pea'' analog, representing a compact starburst with extremely high [O\,\textsc{iii}] equivalent width \citep{cardamone09}. Alternatively, the red-shifted source may be a low-metallicity AGN \citep{groves06}, with the data favoring a scenario where the accreted gas has Z $<$ 0.5Z$_{\odot}$. This scenario is further supported by the observation that the intracluster medium in Abell\,402 has a metallicity of $0.4 \pm 0.2$ Z$_{\odot}$ \citep{accept}. Given the high [O\,\textsc{iii}]/H$\beta$ ratio in the red-shifted source, and the lack of excess blue emission that one would expect from a concentrated starburst, we conclude that the red-shifted source is most likely an accreting SMBH.

\section{Interpretation}

We have shown that the central galaxy in Abell\,402 has a kpc-wide cavity in its stellar mass distribution corresponding to at least $2\times10^9$\,M$_{\odot}$, and a larger deficit of total mass if the dark matter is also missing. Removing this much mass from such a small volume likely requires a dynamical interaction with a scatterer of similar or greater mass. We have also shown that, on larger scales, the central galaxy exhibits a large, diffuse core consistent with a past core-scouring event during the merger and subsequent creation of a $6\pm4 \times 10^{10}$\,M$_{\odot}$ black hole. Finally, we have  shown compelling evidence for the presence of at least one AGN (point-like nature, mid-IR bright power law continuum, LINER-like emission line ratios), and strong evidence for a second AGN (Seyfert-like emission line ratios, point-like to MUSE) with relative velocities of 370\,km/s between the two. As such, we conclude that the stellar cavity is most likely the result of a strong dynamical interaction with the central supermassive black hole and, possibly, with a binary. Below we discuss three possible origins for this cavity.

\subsection{Core Scouring By A SMBH Binary}

A natural explanation for the presence of a kpc-wide cavity in the stellar distribution, representing $>$10$^9$\,M$_{\odot}$ in stellar mass, is three-body scattering of stars and dark matter during the inspiral of two supermassive black holes. Here we investigate this scenario by directly comparing the observations of this system to predictions from two different simulations \citep{merritt06,khonji24}.

Under the assumption that the two emission-line sources are binary SMBHs, and further assuming that they have an orbital plane that is perpendicular to the plane of the sky, we infer a total binary mass via Kepler's law of $M_{tot} = 6 \pm 2 \times 10^{10}$ M$_{\odot}$. If their orbit is inclined into the plane of the sky, this would represent an under-estimate of the binary mass. This high mass is consistent with the measured 2.2\,kpc core (Figure \ref{fig:sbprofile}). Given the large uncertainties on both the progenitor SMBH mass (based on the flat core) and the binary mass (based on the relative velocities of the two AGN), any mass ratio is possible, from an equal-mass merger of two $>$10$^{10}$ M$_{\odot}$ black holes to a highly unequal merger where the primary mass is $\sim$$6\times10^{10}$\,M$_{\odot}$. These masses are generally consistent with observations of $>$10$^{10}$ M$_{\odot}$ black holes at the centers of giant elliptical galaxies \citep{mcconnell11,thomas16,mehrgan19}, but would represent the most massive black hole binary yet discovered by a significant margin. 
We note that the center of this binary is offset both in position ($\sim$0.5\,kpc) and velocity ($\sim$200\,km/s) from the galaxy center and rest frame, respectively, which would indicate that it is not yet in equilibrium. 
The offset binary is consistent with the observation of tidal features in the optical/IR continuum of the central galaxy, and a disturbed X-ray morphology \citep{yuan20}, as one would expect if this galaxy had recently undergone a merger, a necessary condition for the presence of a SMBH binary.

Simulations and analytic theory predict that the separation for a SMBH binary will shrink, or ``harden'', first by dynamical friction from the stars and dark matter, and then by ejection of stars via slingshot interactions, and finally by gravitational radiation \citep{merritt06,khan12,kelley17,sobolenko21,khonji24}. \cite{merritt06} provide a framework for predicting the radiii over which SMBH binary evolution via three-body encounters with stars and dark matter is efficient. This process begins when the separation corresponds to the gravitational influence radius ($r_h$) and will stall when all stars on intersecting orbits have been ejected. The first radius, $r_h$, is well-defined and corresponds to $GM/\sigma^2$. The stalling radius, on the other hand, depends on a variety of factors including the mass ratio of the binary, the total mass of the binary, the velocity dispersion of the stars, and the local slope of the stellar density profile \citep{merritt06}. From the MUSE data, we measure a stellar absorption line velocity dispersion of $375 \pm 15$\,km/s. From the Nuker-profile fits to the F150W2 data (Appendix A.1), we measure an inner slope of $\gamma=-0.02$. Using these values and Table 1 from \cite{merritt06}, we can interpolate a value of $a_{stall}/r_h^{\prime} = 0.06$ for $\gamma=0$ and assuming a 1:1 mass ratio. Using the definition of $r_h^{\prime}$ from \cite{merritt06}, and assuming a total black hole mass of $M_{tot} = 6\times10^{10}$ M$_{\odot}$ and a velocity dispersion of 375\,km/s yields a value of $r_h^{\prime} = 3.7$\,kpc and, thus, a stalling radius of $a_{stall} =  0.18$\,kpc, or a separation of $\sim$0.4\,kpc. This can be compared to the measured separation of $1.2\pm0.3$\,kpc. The fact that the observed separation of 1.2\,kpc lies between the gravitational influence separation (7.4\,kpc) and the stalling separation (0.4\,kpc) implies that this system is in the early stages of binary hardening, when one would expect to observe mass deficits of only $\sim$5--10\% of the binary mass \citep{merritt06}. 

\cite{merritt06} also make predictions for galaxies in the Virgo cluster, which we can directly compare to. The brightest galaxy for which they make this prediction, NGC\,4472, has predicted a stalling radius of $a_{stall} = 5.6$\,pc, for a black hole mass of $5.94 \times 10^8$ \,M$_{\odot}$. Given that $a_{stall} \propto r_h \propto M_{BH}$, we should scale this up by a factor of 100 to match our estimate of the black hole mass in the center of Abell\,402. This yields a stalling radius of 0.56\,kpc, or a stalling separation of 1.1\,kpc, fully consistent with the observed size of the stellar cavity.

\begin{figure}[htb]
\centering
\includegraphics[width=0.99\linewidth]{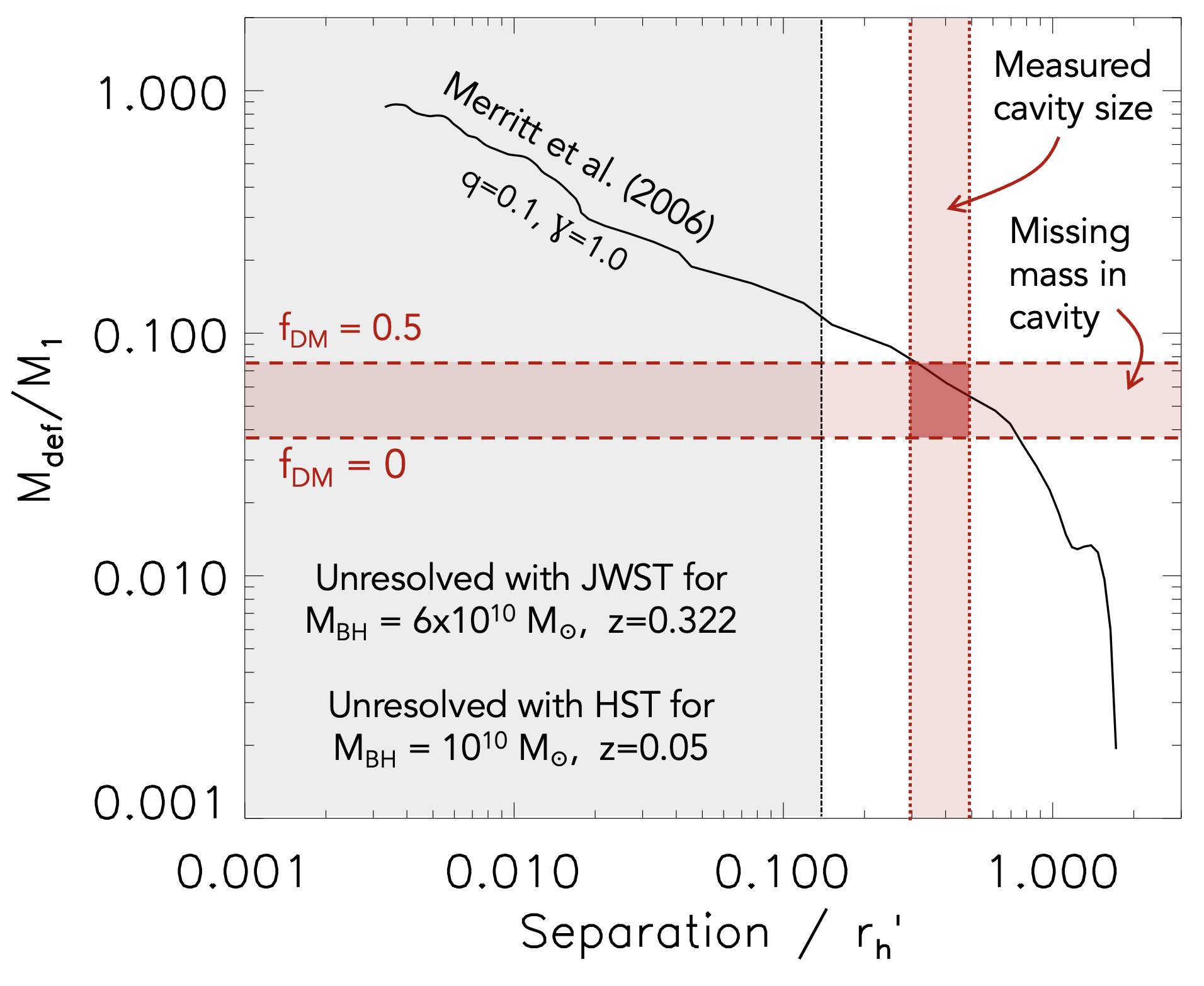}
\caption{An example simulation from \cite{merritt06} showing the amount of mass in stars and dark matter scattered to larger radius during three-body encounters from a binary SMBH as a function of binary separation -- both axes are scaled by a scale factor related to the mass of the binary, making this prediction mass independent. The simulation shown is for a 10:1 ($q=0.1$) binary in a galaxy with a slight stellar cusp ($\gamma=1.0$), which was the only parameter set shown in \cite{merritt06} and is used here for illustrative purposes. While, eventually, the amount of missing mass approaches the total mass of the binary, the missing mass is expected to be closer to 10\% for measurable separations in black holes with $M > 10^{10}$ M$_{\odot}$. For the central galaxy in Abell\,402, we measure a stellar mass deficit of $2\times10^9$\,M$_{\odot}$, which could be a factor of $\sim$2 higher for realistic dark matter fractions in the centers of elliptical galaxies \citep{auger10,barnabe11}. Coupling this with the measured separation ($1.2 \pm 0.3$\,kpc), and normalizing these to the primary SMBH mass (M$_1$) and the gravitational sphere of influence ($r_h^{\prime}$), we find excellent agreement with theory.} 
\label{fig:merrit}
\end{figure}

In Figure \ref{fig:merrit}, we directly compare the measured size and missing mass in the stellar cavity to predictions from simulations \citep{merritt06} to assess whether these observations are consistent with the merging SMBH binary scenario. Given the uncertainties in the dark matter content in the galaxy core and in the measured size of the cavity, we find excellent agreement between the theoretical expectation for a binary SMBH and our observations of a stellar cavity accounting for $\sim$5\% of the SMBH mass. Indeed, theory predicts that this process will only be visible for a short time, before the separation reaches $\sim$10\% of the gravitational sphere of influence and the binary is unresolvable by JWST. This early phase of binary hardening (orbit shrinking from $r_h$ to 0.1$r_h$) should last for $\sim$40\,Myr \citep{khonji24}, and produce $\sim$10--20\% of the mass deficit that will ultimately form the diffuse scoured core \citep{merritt06}. We note that this 40\,Myr timescale is based on simulations which assume that the initial stellar density profile is cuspy -- for a flat core, as we observe in Abell\,402, it may be significantly longer.

Assuming a major merger (M$_1$/M$_2$ $>$ 1/4) rate for the most massive galaxies at $z\sim0.1$ from the Illustris simulations \citep{rodriguez15} of 0.1 Gyr$^{-1}$ and a timescale for which the binary is visible of $\sim$40\,Myr (see above), we would only expect to catch this process in action for $\sim$0.5\% of massive galaxies. This is consistent with the fact that we have found one such system in our larger sample of $\sim$100 clusters. While binary AGN at lower mass should be more common, and have been detected \citep[e.g.,][]{koss12}, the expected size of the scouring region is proportional to the SMBH mass and so the expected angular sizes of such regions would be too small for \emph{HST} or \emph{JWST} to resolve for SMBHs with $M_{BH} < 10^9$\,M$_{\odot}$ outside of $z \sim 0.1$.

These comparisons to simulations are meant to be illustrative and do not, on their own, prove that the central galaxy in Abell\,402 is in the midst of a binary SMBH merger. That said, the simulations do support this interpretation, with the stellar cavity in Abell\,402 being the appropriate size and mass deficit given the observed properties of the central galaxy and an assumed SMBH mass of $\sim$$6\times10^{10}$\,M$_{\odot}$.

\subsection{Post-Merger Black Hole Recoil}

The above interpretations requires the presence of two SMBHs, which we assume to be the two emission line sources. If, instead, there is only one SMBH (the bright source on the western edge of the cavity), then the velocity measured in this source may represent a recoil ``kick'' after the merger of two SMBH \citep[see e.g.,][]{boylan04,favata04,nasim21,khonji24}. 
The magnitude of these velocity kicks are expected to be $\sim$200\,km/s for non-spinning SMBHs \citep{zlochower15}, but can be $\ge$400\,km/s for particular mass ratios and spins \citep{herrmann07}, which is consistent with the velocity offset between the blue-shifted source and the stellar continuum of $\sim$100\,km/s. Given this velocity of the AGN relative to the stars, and a cavity radius of $\sim$0.5\,kpc, the recoil kick would have happened only $\sim$5\,Myr ago (half crossing time of the cavity). Such a short observable timescale makes it highly unlikely (thought not impossible) that we would observe in this state -- as described above, the three-body scouring scenario would be observable for a factor of $\sim$5--10 longer, making it the more likely explanation. Nonetheless, this post-merger scenario is plausible, though it has a low probability of detection and fails to explain the presence of point-like, highly-ionized gas on the eastern edge of the cavity.

Another possible interpretation is that the stellar cavity observed in Abell\,402 may be due to the accumulation of several past scouring and recoil episodes. We note, however, that the observed edges of the cavity are sharp -- when we model it with a S\'ersic function it is best-fit by $n=0.1$ (see Appendix A.1). The sharpness of this feature implies that it cannot be long-lived. Indeed, simulations predict that for a similar-mass galaxy and central SMBH, the central deficit, which is initially quite sharply defined should smear out over several \,kpc post-merger \citep{khonji24}.

\subsection{A Dipole Instability in the Stellar Distribution}

\cite{dattathri25a} provide the framework for an alternative explanation for the observed cavity in the center of Abell\,402-BCG.  Stellar distributions with a rapid transition between the inner and outer slopes can violate Antonov's stability criterion. In particular, simulations predict that such systems should have instabilities that manifest as a rotating dipole in the stellar density distribution. The instability arises from an inflection in the system's isotropic distribution function, which characteristically appears in systems with large, flat cores \citep[see Figure 2 in][]{dattathri25a}. For the central galaxy in Abell\,402, which has which, based on a double-powerlaw fit, has an inner slope of $\gamma=0.14$ and outer slope $\alpha=4.13$ (see Figure \ref{fig:sbprofile}), we expect the system to go unstable interior to the transition radius at $\sim$2.2\,kpc.

Based on Figure 11 from \cite{vandenbosch25}, the dipole instability is expected to produce positive and negative deviations of the order tens of percent from the average, with the fluctuations having similar size to the cavity observed in Abell\,402.  So, at least qualitatively, the properties of the cavity are consistent with a stellar dipole. The production of this dipole can torque a central SMBH outward, leading to off-center black holes in galaxy cores. It can also act as a source of buoyancy that slows down the inspiral of additional black holes, delaying their mergers \citep{dattathri25b}. In the case of Abell\,402-BCG, the scouring caused during the formation of the ultramassive central black hole may have caused the core to go dipole unstable. This in turn started a complicated dynamical interaction between the dipole and the two SMBHs, leading to the observed configuration. The fact that the brightest and most secure AGN seems to be located close to the peak in the central surface brightness is consistent with the simulations of \citet{dattathri25b}.

While this explanation for the stellar cavity is appealing as it seems to be a natural consequence of a flat stellar core, it also faces some challenges. Most critically, the central structure of Abell\,402 does not show as clean a dipolar structure with a positive fluctuation on the opposite side, as seen in the simulations of \citet{dattathri25a}. In particular, in Appendix A.1, we show that the galaxy in all five filters can be modeled as the combination of a symmetric Nuker model, a negative fluctuation, and a point source, with no strong residuals. However, one could imagine that a combination of projection and general asymmetry, combined with the point source overlapping with the expected location of the positive fluctuation, could mask some of this. In addition, simulations of this dipole instability \citep{dattathri25a,dattathri25b} have only been considered for symmetric systems with a single black hole -- it may be that more complicated systems with either a recoiling SMBH or a binary SMBH could interact with the dipole in such a way to produce the observed signature in Abell\,402.  Nonetheless, this mechanism is intriguing as it suggests that a strong, localized, negative fluctuation in the stellar density field, as observed in Abell\,402, may be a consequence of an instability of the core region, rather than of dynamical interactions with a binary SMBH. However, we emphasize that the creation of the (unstable) core itself would still likely have been a consequence of recent scouring or recoil scenarios.

\section{Conclusions}


We report the discovery of a kpc-wide dark patch in the center of Abell\,402 BCG based on data from \emph{HST} and \emph{JWST}. We have shown that this feature is not caused by dust absorption and, instead, appears to be a cavity in the stellar distribution that represents a missing mass of $>$$2\times10^9$\,M$_{\odot}$. In addition to this ``stellar cavity'', we find that the central galaxy in Abell\,402 has an extremely large diffuse core, with a break radius of 2.2\,kpc, on which the (smaller) cavity is superimposed. Such a large core was likely produced during the past merger of SMBHs, leaving behind a remnant with a mass of $6\pm4\times10^{10}$\,M$_{\odot}$. This ``ultra-massive'' black hole appears on the western edge of the stellar cavity as a mid-IR bright point source in the \emph{HST} and \emph{JWST} data, and is coincident with a LINER AGN as identified by MUSE. On the eastern edge of the cavity we find evidence for a second AGN, based on the presence of strong, localized [O\,\textsc{iii}] emission -- these two AGN have a relative velocity of 370\,km/s, implying a combined binary mass of $6\pm2\times10^{10}$\,M$_{\odot}$. 

We postulate that the stellar cavity is being formed by the ongoing dynamical interaction of a stellar core with in-spiraling SMBHs. This interpretation is supported by the evidence in the optical and X-ray for a recent galaxy-scale merger and by the observation of two emission line sources on either side of the cavity, consistent with a massive binary AGN.  The size of the cavity and the relative velocities of the two purported AGN are consistent with decades of theory on three-body scouring of stellar cores by in-spiraling SMBHs. Other valid hypotheses for the origin of this cavity include (i) a post-merger recoiling SMBH, which would lead to the rapid expansion of stellar orbits on small scales, or (ii) the development of a dipole instability due to the rapid change in slope of the stellar surface density profile. 

This system provides a blueprint for a new phenomenon to look for in existing and future observations -- multi wavelength data was essential in distinguishing between dust absorption (wavelength-dependent) and missing stars (wavelength-independent).  
Detection of other similar sources, or confirmation of the binary AGN hypothesis in this system, would help firm up our current predictions of multi-messenger signatures for individual LISA sources, for which the merger timescale and occurrence remain key uncertainties.
Moving forward, large surveys with \emph{Euclid} and  \emph{Roman}, as well as archival surveys through the \emph{HST} and \emph{JWST} archives, will likely uncover similar systems, providing an estimate of the SMBH merger timescale based on the frequency with which such systems are observed. At the same time, targeted follow-up of this unique system with our most powerful telescopes will facilitate a more complete understanding of the nature of the AGN in this system and the dynamics of the stars in and around the cavity.


%

\begin{acknowledgments}

We thank the anonymous referee for thoughtful and helpful comments that led to significant improvements in this paper.
This work is based in part on observations made with the NASA/ESA/CSA \emph{JWST}. The data were obtained from the Mikulski Archive for Space Telescopes at the Space Telescope Science Institute, which is operated by the Association of Universities for Research in Astronomy, Inc., under NASA contract NAS 5-03127 for \emph{JWST}. These observations are associated with \emph{JWST} Cycle 1 GO program 5594. Support for the program JWST-GO-5594 was provided by NASA through a grant from the Space Telescope Science Institute, which is operated by the Associations of Universities for Research in Astronomy, Incorporated, under NASA contract NAS5-26555. This research is based on observations made with the NASA/ESA Hubble Space Telescope obtained from the Space Telescope Science Institute, which is operated by the Association of Universities for Research in Astronomy, Inc., under NASA contract NAS 5–26555. These observations are associated with programs HST-GO-14148, HST-GO-10875, HST-GO-12166, HST-GO-16262, and HST-GO-6554.
This research was supported in part by the University of Pittsburgh Center for Research Computing, RRID:SCR\_022735, through the resources provided. Specifically, this work used the H2P/MPI cluster, which is supported by NSF award number OAC-2117681. 
M.\ McDonald acknowledges support from JWST-GO-05594.013-A, which was provided through a grant from the STScI under NASA contract NAS5-03127.
J.\ Diego  acknowledges support from project PID2022-138896NB-C51 (MCIU/ AEI/ MINECO/ FEDER, UE) Ministerio de Ciencia, Investigaci\'{o}n y Universidades.
A.\ C.\ Edge and R.\ Massey are supported by UK STFC (grant ST/X001075/1).
M.\ Jauzac and S.\ Werner are supported by the United Kingdom Research and Innovation (UKRI) Future Leaders Fellowship ``Using Cosmic Beasts to uncover the Nature of Dark Matter'' (grant number MR/X006069/1).
M.\ Montes acknowledges support from grant RYC2022-036949-I financed by the MICIU/AEI/10.13039/501100011033 and by ESF+ and program Unidad de Excelencia Mar\'{i}a de Maeztu CEX2020-001058-M.
P.\ Natarajan acknowledges support from the Gordon and Betty Moore Foundation and the John Templeton Foundation that fund the Black Hole Initiative (BHI) at Harvard University where she serves as one of the PIs.
%
 %
A.\ Zitrin acknowledges support by grant 2020750 from the United States-Israel Binational Science Foundation (BSF) and grant 2109066 from the United States National Science Foundation (NSF), and by the Israel Science Foundation Grant No. 864/23.

\end{acknowledgments}





%
\facilities{HST(ACS, WFC3), JWST(NIRCam), Chandra, ALMA, EVLA, VLT(MUSE)}




\appendix

\renewcommand\thefigure{\thesection.\arabic{figure}}    
\setcounter{figure}{0}  

\section{Appendix}

\subsection{Two-Dimensional Modeling of HST/JWST Images}
In Figure \ref{fig:fig1}, we show 5-band imaging of the central galaxy in Abell\,402, taken with cameras on \emph{HST} (F606W, F814W, F110W) and \emph{JWST} (F150W2, F322W2). These images depict three clear morphological components: (i) an elliptical galaxy, (ii) a ``cavity'' on the left side of the galaxy, which is more obvious in the F606W, F814W, and F150W2 bands than in the F110W and F322W2 bands, and (iii) a point source, which is most obvious in the F322W2 and F150W2 bands, and barely visible in the F606W band. Each of these bands has different pixel scales and different point spread functions, which makes it non-trivial to assess whether the observed cavity and/or the point source are varying with bandpass.

To assess the wavelength dependence of these morphological features, we have modeled the central region of Abell\,402 in all five bands using \textsc{Sherpa}\footnote{https://cxc.cfa.harvard.edu/sherpa/}. We note that, while \textsc{Galfit} \citep{galfit} is more commonly used to model surface photometry of galaxies, it does not allow for negative components (given that it works in magnitude units), which would make it impossible to add a ``cavity'' component.
We use the combination of a a two-dimensional Nuker profile \citep{lauer95,graham03} to describe the central galaxy, a two-dimensional S\'ersic profile with a negative amplitude to describe the cavity, and a point-like Gaussian to describe the point source. In the F150W2 band, all structural parameters were allowed to vary freely, including the inner and outer slopes of the Nuker profile, the size and sharpness of the cavity, and the location of the point source. The F150W2 band was fit first, with the best-fit parameters used as inputs for the other four bands. To properly sample our modeling uncertainty, we fit each image 50 times. We used 4 point sources in the field of view as our template PSF, randomly choosing one for each bootstrap fitting and allowing for a 0.1-pixel centering uncertainty for this PSF model. We also, for each bootstrap, randomly sample the intensity at each pixel based on the local RMS. 

\begin{figure}[h!]
\centering
\includegraphics[width=0.99\linewidth]{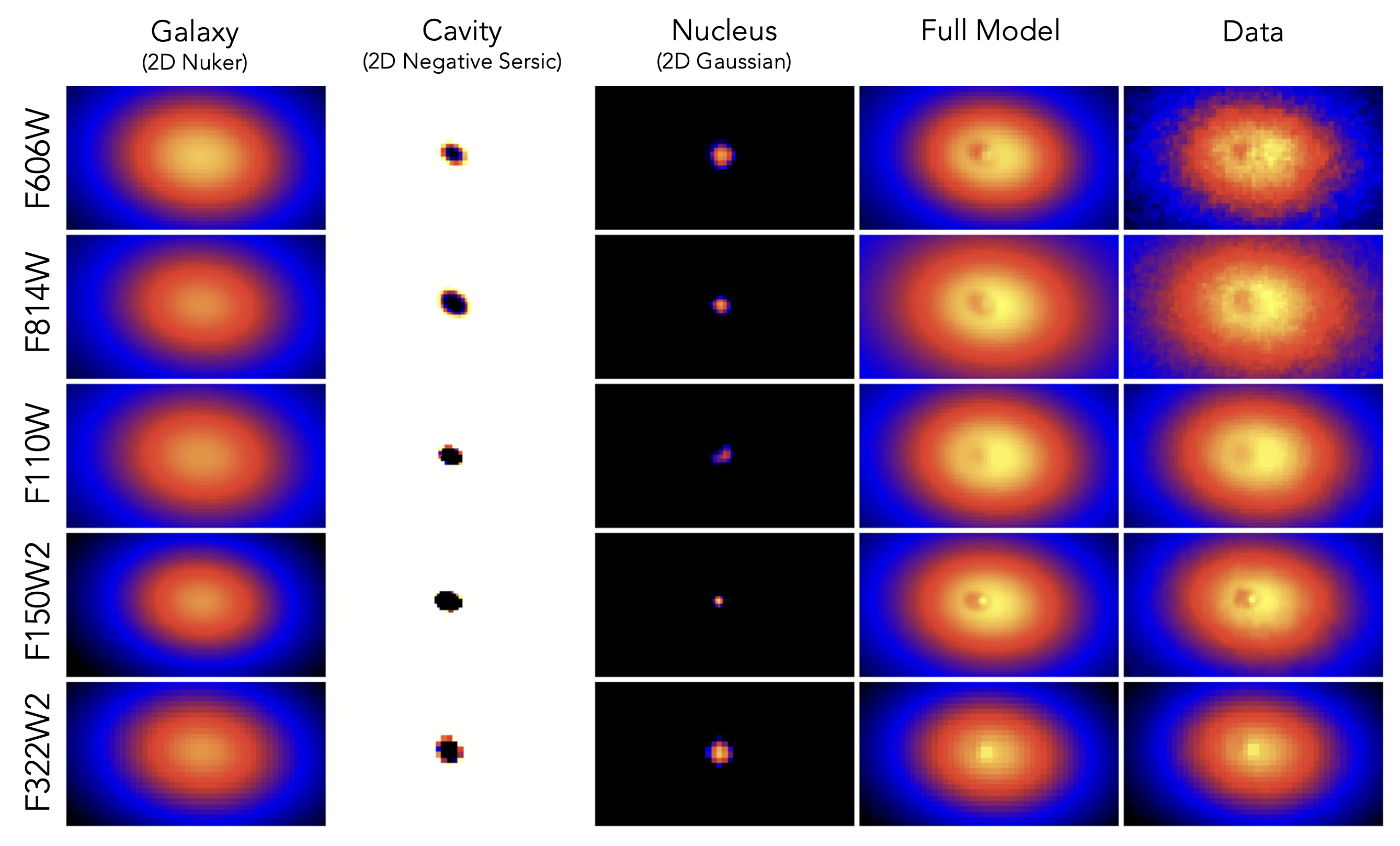}
\vspace{-0.1in}
\caption{The surface photometry for the central galaxy in Abell\,402 has been modeled using a combination of a Nuker profile \citep{graham03} for the ``galaxy'', a negative-intensity S\'ersic profile for the ``cavity'', and a point-like Gaussian for the ``nucleus'', all convolved with the instrumental PSF. All models shown represent the average model fit when marginalizing over variations in the PSF and the count rate. The component models are shown without the PSF convolution applied, whole the ``full model'' panels include the instrumental PSF. This figure demonstrates that the large variations in the observations can be attributed to a central AGN with a steep wavelength dependence and a PSF that is a factor of two larger at F110W and F322W2 compared to F606W, F814W, and F150W2. This image decomposition allows us to accurately measure the total flux missing in the cavity (Figure \ref{fig:fig2}), the spectral energy distribution of the missing stars in the cavity and the galaxy (Figure \ref{fig:supp2}), and the spectral energy distribution of the point source (Figure \ref{fig:supp3}).}
\label{fig:supp1}
\vspace{-0.1in}
\end{figure}

The results of this analysis are shown in Figure \ref{fig:supp1}. For the models, we show the average of 50 bootstrapped fits to the data, accounting for small errors in the PSF and source counts between fits. The best-fit models are very similar between filters, despite the fact that all parameters were allowed to vary. These models allow us to determine the amount of flux missing in the cavity as a function of wavelength, by simply integrating over the ``cavity'' model -- these ``missing flux'' values are presented in Figure \ref{fig:fig2} -- this will be discussed further in the following section. 

From Figure \ref{fig:supp1} we can also assess the color gradient in the underlying galaxy light profile. We use these color gradients to create an empirical model for the expected amount of missing flux due to a cavity in the stellar distribution as a function of wavelength. More specifically, to predict the effects of a gap in the stellar distribution, we compare the intensity in a given band at two galaxy radii, where the intensity at larger radius can be thought of as containing all of the same line-of-sight contributions as that at smaller radius, but with a segment missing from the mid-plane with length corresponding to the difference in radii. In this way, we use the color gradients to model a cylindrical hole with a depth of 1\,kpc and a separation of 1\,kpc from the center of the galaxy's light distribution in the plane of the sky. This model is shown as a grey band in Figure \ref{fig:fig2}, where the uncertainty corresponds to our uncertainty in the 2-D position of the cavity in a given filter. This figure confirms our intuition that a gap in the stellar distribution would produce a flat profile in the amount of missing flux as a function of wavelength. If the galaxy had a strong color gradient (which the central galaxy in Abell\,402 does not have) then we would expect a slight slope in this line.

\subsection{Spectral Energy Distribution Modeling}

\begin{figure}[h!]
\centering
\includegraphics[width=0.6\linewidth]{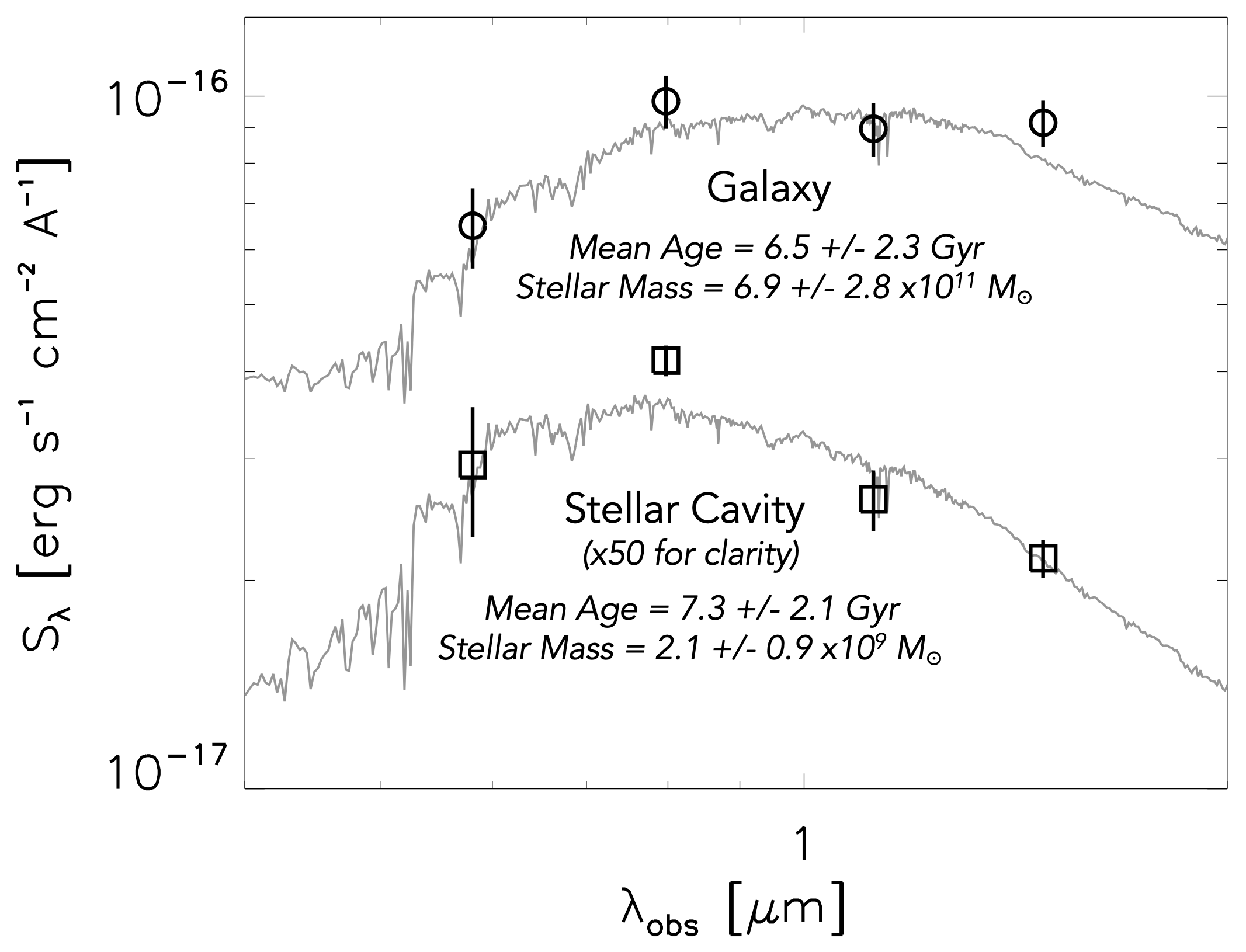}
\vspace{-0.0in}
\caption{This figure shows the integrated model photometry for the central galaxy and the stellar cavity in Abell\,402. For clarity, the flux values for the stellar cavity have been scaled by a multiplicative factor of -50 to make them positive and similar in intensity to the galaxy. The best-fit models from CIGALE \citep{cigale1,cigale2} are shown in grey, and the relevant properties of these stellar populations are included. This figure demonstrates that the stellar cavity has a spectral energy distribution consistent with missing stars, and that the implied stellar mass that is missing is $\sim 2\times10^9$\,M$_{\odot}$.}
\label{fig:supp2}
\end{figure}

To estimate the missing stellar mass in the kpc-wide cavity, we utilize the image decomposition described in the previous section, which yields an analytic model for the cavity as a function of wavelength. We sum up the total missing flux in this component of the model for each filter, and perform flux calibration using the relevant header information in the data. For comparison, we also sum up all of the light in the Nuker model component for each filter. The resulting spectral energy distributions (SEDs) are shown in Figure \ref{fig:supp2}, where we have excluded the F322W2 data here due to the fact that it is the most affected by PSF modeling. We fit these two SEDs using PCIGALE \citep{cigale1,cigale2}, using a delayed star formation history with an optional burst, a Salpeter IMF, and minor contributions from dust, nebular lines, and residual contamination from the AGN. The best-fit models for both the underlying galaxy and the missing stellar mass are shown in grey on Figure \ref{fig:supp2}. From this modeling, we can deduce the total stellar mass that is contained in this negative-intensity component, finding $M_{*,cavity} = 2.1 \pm 0.9 \times 10^9$ M$_{\odot}$. This represents $<$1\% of the total stellar mass of the galaxy. The SED of the missing mass is fully consistent with that of the underlying galaxy, as is depicted in Figure \ref{fig:supp2}, with best-fit ages of $6.5\pm2.3$\,Gyr and $7.3\pm2.1$\,Gyr.

\begin{figure}[h]
\centering
\includegraphics[width=0.8\linewidth]{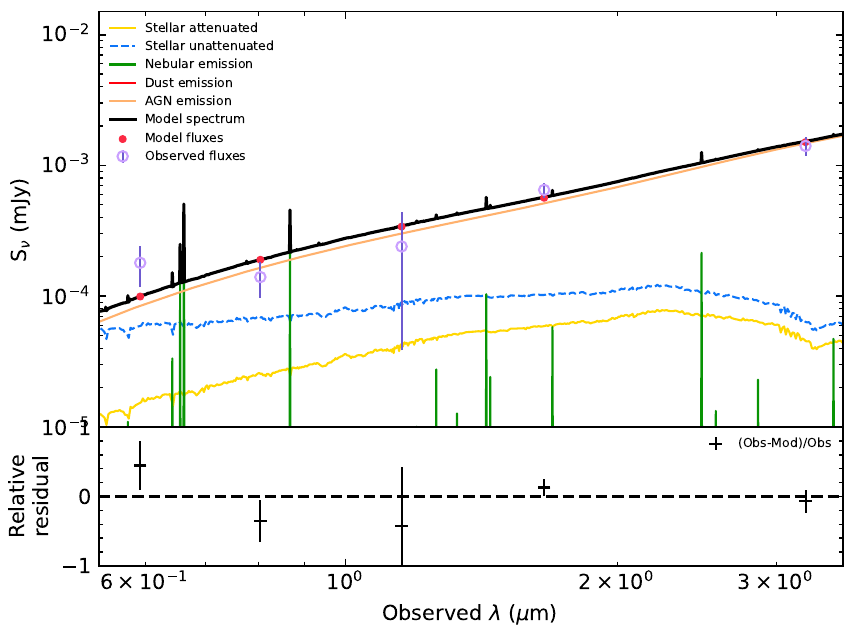}
\vspace{-0.0in}
\caption{This figure, produce by the PCIGALE software \citep{cigale1,cigale2}, shows the fluxes obtained for the central point source in each filter based on the two-dimensional image decomposition described in previous sections. This spectral energy distribution is well-described by a power law at all wavelengths. We model these data using PCIGALE, allowing for the combination of a two-stage star formation history, nebular emission, and an AGN, finding that the emission is best described by an AGN with contributions from nebular lines at the shortest wavelengths. The fact that these data are well-described by a single power law, despite being obtained over the course of 20 years, suggests minimal time variability in the central point source.}
\label{fig:supp3}
\end{figure}

We additionally model the broadband spectral energy distribution of the central point source, to determine if it is consistent with an AGN and whether or not time variability is required to explain the SED. From the two-dimensional image decomposition described above, we obtain fluxes and their model uncertainties of the central point source in each filter, which are shown in Figure \ref{fig:supp3}. We again use PCIGALE \citep{cigale1,cigale2} to model this SED with a combination of a stellar component and an AGN component, finding that the data is well-described by an AGN (powerlaw and nebular lines) at all wavelengths, with a reduced $\chi^2 = 1.1$. The most discrepant band is F606W, which lies $\sim$1.5$\sigma$ above the model. This offset may be due to time variability (the F606W observations were taken 18 years before the JWST observations), due to an emerging contribution at short wavelengths from a nuclear starburst, or may be a statistical fluctuation. Given the significant uncertainties in the modeling, we are unable to distinguish between these cases. We note that a follow-up observation at any of these wavelengths would establish whether or not time variability is significant, but conclude that the evidence for this is statistically weak with the data in hand.

\subsection{Additional Multi-Wavelength Data}

When evaluating the binary SMBH hypothesis, we considered all publicly available radio, mm-wave, and X-ray data. Generally speaking, these data are challenging to compare to the JWST/HST data, due to the fact that the angular resolution of the optical/IR data is more than an order of magnitude better than the best radio/X-ray data available. However, we can still learn something from these comparisons, as we outline below.

Radio data is available for the central galaxy in Abell\,402 from a variety of telescopes. The highest angular resolution data available is from EVLA -- the VLA Sky Survey covers this field at 1.5\,GHz with an angular resolution of 2.5$^{\prime\prime}$ and pointed observations in the B configuration at 6\,GHz (Program AE223; PI: A.\ Edge) yield an angular resolution of 1.0$^{\prime\prime}$. Neither of these observations yield a detection at the location of the central galaxy in Abell\,402, implying an upper limit of $\sim$0.1mJy. At lower frequencies, there is a tentative detection of \emph{extended} emission at 300\,MHz from the GMRT \citep{cuciti21}, which the authors suggest may be a radio mini halo, but caution that it is only a ``probable'' detection. From these data, we can conclude that there are no strong radio jets in this system, which suggests that the double-peaked velocity for the emission lines in the optical is not related to a radio-powered outflow.

The ALMA observatory has also pointed at Abell\,402 in several configurations. Unfortunately, the three pointings with high angular resolution ($<$0.2$^{\prime\prime}$) are offset from the central galaxy, presumably targeting a lensed background galaxy. These data are not helpful for understanding the nuclear region of the central galaxy. The remaining two observations have angular resolutions of 0.5$^{\prime\prime}$ and 1.0$^{\prime\prime}$, targeting frequencies around 95\,GHz and 140\,GHz, respectively. While the frequency bands chosen for these data do not overlap with any strong CO transitions at $z=0.322$, we still examined them both channel by channel, finding no evidence for line emission in the galaxy center. In addition, the continuum maps contained no significant emission coincident with the galaxy, suggesting that neither of the line-emitting sources in the optical are bright in mm-wave continuum.

\begin{figure}[h!]
\centering
\includegraphics[width=0.9\linewidth]{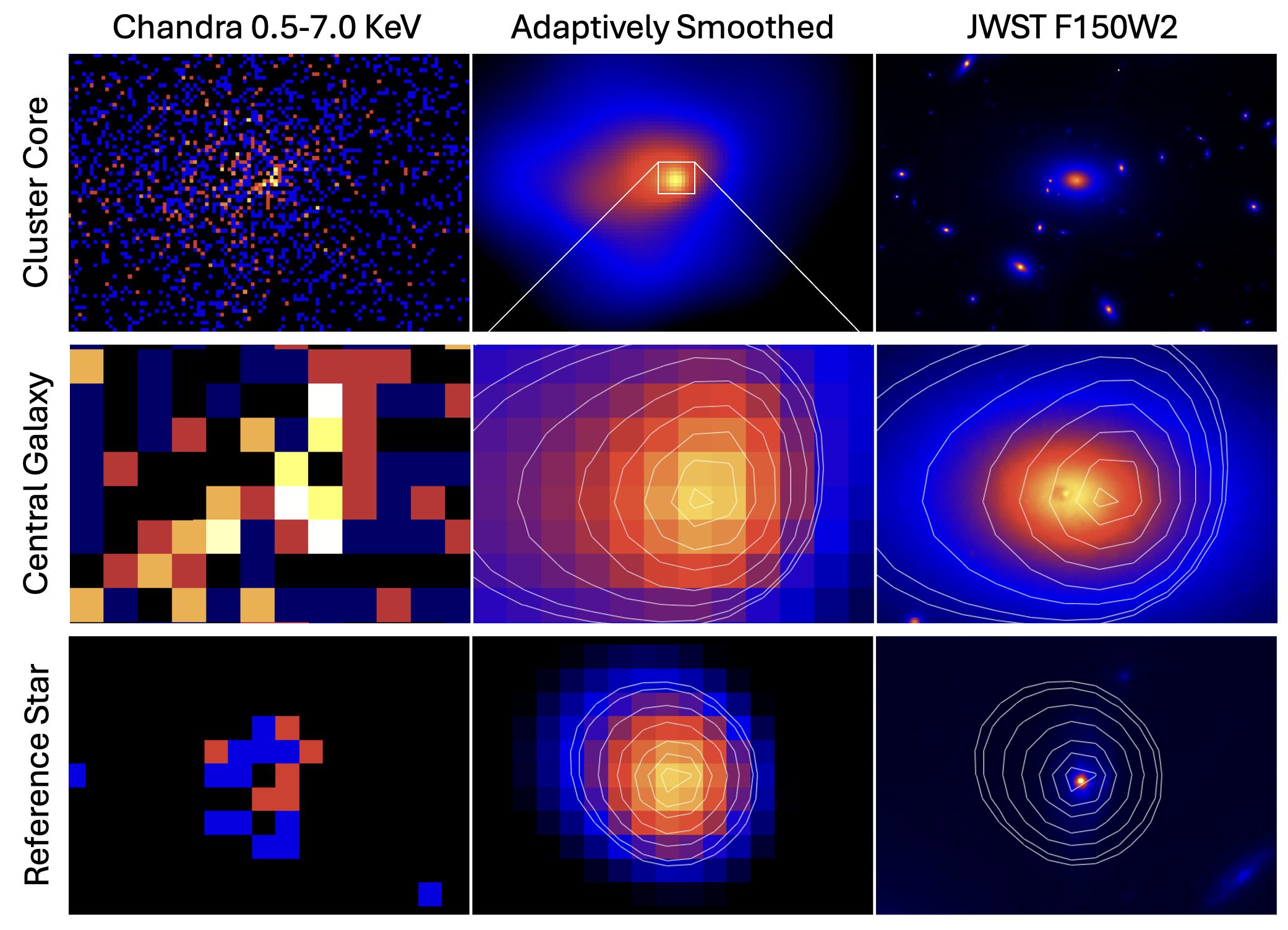}
\vspace{-0.1in}
\caption{\emph{Chandra} observations of Abell\,402. \emph{Upper Panels:} Large-scale X-ray emission, which exhibits a clearly asymmetric morphology indicative of a recent cluster-scale merger. \emph{Middle Panels:} Zoom in on the X-ray peak, which is significantly offset from the galaxy center and the IR-bright point source. We interpret this as a sloshing core, rather than a point source, suggestive of a recent merger. \emph{Bottom Panel:} Zoom in on a nearby point source, to emphasize the excellent astrometry of these data, indicating that the offset above is not an astrometry error.}
\label{fig:supp_chandra}
\vspace{-0.1in}
\end{figure}

Finally, high angular resolution X-ray data (20\,ks; OBSID 3267) is available for this source from \emph{Chandra}. We have downloaded and reprocessed these data, extracting an image in the 0.5--7.0\,keV bandpass. The counts images have been adaptively smoothed to a pixel-level signal-to-noise ratio of 2.5, with a minimum smoothing scale of 0.5 pixels. The resulting images are shown in Figure \ref{fig:supp_chandra}. On large scales, it is apparent that the X-ray surface brightness is sharply peaked, and that the cluster has a generally asymmetric morphology, with an elongated morphology that is often seen in clusters that have recently experienced a merger. This is consistent with published works labeling this cluster as unrelaxed \citep{yuan20}. Zooming in on the central X-ray peak (middle panels), we see that it is not coincident with the IR-bright point source, which lies outside of the FWHM of the Chandra PSF. Furthermore, this point source is brightest in the soft X-rays, and is undetected in a 2--8\,keV bandpass, suggesting that it is not an AGN. While the presence of this point-like source is enticing, and would make sense given the AGN-like nature of the IR-bright point source, we interpret this morphology as a ``sloshing core'', which are indicative of a recent merger with another massive structure \citep{zuhone11}. For comparison, we show in the lower panels of Figure \ref{fig:supp_chandra} the same data for an X-ray point source that lies 1.5$^{\prime}$ from the center of the cluster and is in both the \emph{Chandra} and \emph{JWST} fields of view. This X-ray point source, which is also present in a hard X-ray band, is well-aligned with a near-IR point source, suggesting that the offset observed in the center of the galaxy is not an astrometry error.

In summary, the available multi-wavelength data can rule out that the two line-emitting sources are not radio jet hot spots, and that the cavity is not caused by a large cloud of molecular gas. They do not, however, provide any additional supporting evidence for the dual-AGN hypothesis.


\bibliography{ref.bib}{}
\bibliographystyle{aasjournalv7}



\end{document}